# Electron Attachment to Wobble Base Pairs

*Jishnu Narayanan S J, Arnab Bachhar, Divya Tripathi, and Achintya Kumar Dutta\**

*Department of Chemistry, Indian Institute of Technology Bombay, Powai, Mumbai-400076*

**Abstract**

We have analyzed the low-energy electron attachment to wobble base pairs using the equation motion coupled cluster method and extended basis sets. A doorway mechanism exists for the attachment of the additional electron to the base pairs, where the initially formed dipole-bound anion captures the incoming electron. The doorway dipole-bound anionic state subsequently leads to the formation of a valence-bound state, and the transfer of extra electron occurs by mixing of electronic and nuclear degrees of freedom. The formation of the valence-bound anion is associated with proton transfer in hypoxanthine-cytosine and hypoxanthine-adenine base pairs, which happens through a concerted electron-proton transfer process. The calculated rate constant for the dipole-bound to valence-bound transition in wobble base pairs is slower than that observed in the Watson-Crick guanine-cytosine base pair.

\*achintya@chem.iitb.ac.in

**Introduction**

The secondary electrons (SE) are one of the major contributors to the radiation damage to genetic material. They are generated by the interaction of high-energy ionizing radiation with the cellular medium. Although $\sim 4\times10^4$ SE are produced per MeV of radiation-energy supplied, through ionization of water,[1] a significant portion of the electrons lose their energy through solvent-electron interaction and form solvated electrons $(e_{aq})$.[2] The conversion of SE to $e_{aq}$ proceeds through the formation of low energy electrons (LEE), which have energies less than 20 eV. These LEEs can cause damage to genetic material via dissociative electron attachment (DEA).[3–5] It was earlier believed that direct radiation-DNA interaction resulted in DNA strand breaks and, ultimately, apoptosis of cancerous cells during radiation cancer therapy. However, Woldhuis et al.[6] and Sanche and co-workers[7] provided experimental evidence of electron attachment-induced radiation damage to genetic material in their seminal works.

Many theoretical and experimental works have been conducted to unravel the intricacies of the process of LEE attachment to DNA and its subsequent damage.[7–24] Since DNA is a macromolecule, the entire DNA or RNA strand cannot be simulated using the existing quantum chemical techniques due to computational bottlenecks. Therefore, we often build appropriate model systems, such as nucleobases, base pairs, nucleotides, etc., to describe electron attachment to DNA. Among the various model systems employed to simulate electron attachment to DNA *in silico*, a nucleobase pair may be regarded as one of the simplest but chemically sound ones. The base pair model also preserves the strong inter-base interactions within the double helix, which plays a significant role in intra-DNA charge transfer processes caused by LEE attachment. Several types of base pairing are observed in biomolecules, which can be broadly classified into two categories: canonical and non-canonical base pairs. Watson-Crick (WC) adenine-thymine (A:T) and guanine-cytosine (G:C) are the canonical base pairs and contribute to almost all of the base-pairing interaction found in the DNA double helix. Non-canonical base pairs can include Hoogsteen, wobble, reverse-WC base pairs, etc. LEE attachment to neutral base pairs leads to the formation of radical anions. For convenience, radical anions will be referred to as anions from hereafter unless specified otherwise.

One of the first theoretical studies on the electron attachment to WC base pairs was done by Sevilla and co-workers at the Hartree-Fock (HF) level of theory.[25] According to their results, a valence-type anion of A:T and G:C base pairs are formed upon electron attachment, which also leads to proton transfer between the nucleobases. Adamowicz and co-workers employed second-order Møller–Plesset perturbation theory along with Pople 6-31+G(d,p) basis set to suggest that LEE attachment to A:T base pair is not favorable.[26] It was because they obtained a negative value for the adiabatic electron affinity (-0.40 eV) for A:T. They further extended their calculation to WC G:C base pair.[27] Contrary to the previous HF results, they predicted that the dipole-bound state of G:C anion exists and can coexist with the valence-bound anion where the additional electron density is localized on cytosine. Their MP2/6-31++G**(6d) results for adiabatic electron affinity (AEA) of G:C shows an increase of 0.29 eV compared to that calculated at the HF level. However, MP2 cannot be considered an ideal theory to simulate electron attachment to genetic materials because it is prone to issues such as symmetry breaking and spin contamination in the case of anionic systems.[28] Density functional theory (DFT) has been widely used in the literature to model dissociative electron attachment to DNA. The main advantages of DFT are its favorable scaling and faster convergence with respect to the basis set size, the lack of which are the traditional problems of wave-function-based methods. Schaefer and co-workers used a hybrid-GGA functional B3LYP and TZ2P++ basis set to study the WC G:C anion.[29] They have analyzed the changes in the natural population and the geometry when the anion is formed by electron attachment to the neutral G:C base pair. Cytosine exhibited larger geometrical distortions than guanine in the G:C anion compared to the neutral. Hence, they concluded that the additional electron is localized on the pyrimidine nucleobase. They have also explored the electron attachment to the WC A:T base pair at the same level of theory.[30] In contradiction to the MP2 results of Adamowicz and co-workers, they reported a stable A:T anion with AEA of 0.31 eV. Although DFT provides reliable results when the ground state of a neutral system is considered, it performs poorly for the anions due to the issue of self-interaction.[31] Wavefunction-based methods are free from the problem and offer a more accurate alternative to DFT, albeit at the expense of higher computational cost. [32]

The electron affinities in DFT and MP2 are mostly determined using the Δ-based approach, where the electron affinity is calculated as the difference between the energies of the neutral and anionic species. Hence, such methods require two different calculations to be performed. However, many of the anionic states are not bound at the Hartree-Fock level, and the electron correlation make them bound. When extended basis sets are used for simulating such correlation-bound anions, variational collapse of the anion wave function can occur. Voora et al. have shown that even a CCSD(T) level treatment of electron correlation cannot rectify the variational collapse in the HF wave function.[33] Whereas direct energy difference methods, where the electron affinity is determined as a transition from neutral to anionic state, are free from the problem of the variational collapse of the reference wave function. In this work, we have used the equation of motion coupled cluster (EOM-CC)[34–36] method, which is one of the most accurate among the various direct energy difference-based approaches. At the same time, the accuracy of EOM-CC is also systematically improvable. It is generally used in singles and doubles approximation (EOM-CCSD) and has proved to be an accurate tool for calculating electron affinity values of molecules and clusters. However, performing canonical coupled-cluster calculations on molecules bigger than a single nucleobase is not practical due to their high computational cost. The recent implementation of the domain-based pair natural orbital (DLPNO) formulation of EOM-CCSD (EOM-DLPNO-CCSD)[37] rectifies these issues to a large extent. Electron affinities of large molecules and clusters can be calculated using EOM-DLPNO-CCSD with systematically controllable accuracy.[38] The EOM-DLPNO-CCSD method has been successfully used to understand the electron attachment pathways in WC G:C and A:T base pairs.[39,40]

One can clearly see that most of the theoretical studies on electron attachment to genetic materials using base pairs as model systems have concentrated on Watson-Crick (WC) base pairs (canonical base pair).[25–27,29,30,41] However, to get a complete picture, one needs to consider the non-canonical base pairs as well. Wobble base pairs constitute a significant portion of the total base-pairing interaction in RNAs.[42,43] They are predominantly involved in stabilizing the tertiary structure of RNA, which is essential for their catalytic activity.[43] RNAs showcase a wide variety of complexes with metal ions, proteins, ligands, etc., where wobble base pairs play a significant role. Wobble base pairs, especially Guanine-Uracil (G:U), are fundamental in the codon-anticodon interaction in tRNA during translation in the central dogma.[44] Hence, as it is for WC base pairs, it is essential to probe the mechanism of LEE attachment to non-canonical base pairs as well. In this work, we study the electron attachment to wobble guanine-adenine (G:U), hypoxanthine-cytosine (I:C) , hypoxanthine-uracil (I:U), and hypoxanthine-adenine (I:A) base pairs.

**Theory and Computational Details**

The geometries of the non-canonical base pairs were optimized using the second-order Møller-Plesset perturbation theory under RI approximation[45,46] and def2-TZVP[47] basis set. Vibrational frequency analysis was also done at the same level of theory. For calculating the vertical electron affinities (VEA), we have used the EA-EOM-DLPNO-CCSD method as implemented in ORCA[48–50] along with Dunning's[51] correlation consistent aug-cc-pVXZ(+5s5p4d) (X = D, T) basis set. Additional five s-type, five p-type, and four d-type diffuse functions were added to the standard aug-cc-pVXZ set in an even-tempered way on the atom located closest to the positive end of the dipole-moment vector to simulate the dipole-bound anionic state accurately. TightPNO settings were used for these calculations. Nudged elastic band (NEB)[52] calculations were done to identify the minimum energy path (MEP) for the proton transfer observed in anions of I:C and I:A at RI-MP2/def2-TZVP level of theory.

**Result and Discussion**

Electron attachment to a neutral molecule is generally followed by geometric reorganization as the system tries to minimize the electron-electron repulsive interactions. All four Wobble base pairs form a dipole-bound bound state upon initial electron attachment. The additional electron is bound to the base pair via charge-dipole interaction in the doorway state. As a result, the additional-electron density in the dipole-bound anion is located away from the geometric framework (figure 1). Whereas in a valence-bound anion, the extra electron occupies the antibonding molecular orbital of the neutral base pair. If the additional electron density is localized on the nuclear framework, one should expect considerable changes in the molecular geometry of the anion compared to the neutral molecule. On the other hand, if it is located away from the molecule, the extra electron essentially does not offer significant repulsion to the existing electrons in the system, and hence, minimal geometric distortion will occur in the anionic system. Therefore, the geometry of the valence-bound anion is visibly different from the neutral one since extra-electron density is localized on the system. However, as previously shown, the dipole-bound anionic geometry is almost identical to the neutral geometry.[53,54] Therefore, we have considered the geometry of the dipole-bound anion to be the same as that of the neutral in this work. In WC base pairs, the molecular orbital (MO) corresponding to the valence-bound state shows that pyrimidines have a larger affinity for the extra electron, when paired with their complementary bases.[39,54] We have also made a similar observation with the

Wobble base pairs (figure 1), where the additional electron density in G:U, I:C, and I:U are also localized on the pyrimidine bases. In the case of the I:A base pair, the electron density is localized on the adenine.

Among all the four base pairs, the RMSD values (Table 1) show that the geometry of the valence-bound anion of G:U has the most deviation compared to the corresponding neutral analog. From Figure 2, it can be seen that Guanine (G) and Uracil (U) lose their coplanarity, and as a result, the hydrogen bond between $O_4$ of guanine and $H_{28}$ of uracil gets broken once the valence-bound anion is formed. The excess negative charge, localized on U in G:U valence-bound anion, enhances the hydrogen bond acceptor character of $O_{17}$ of U. In addition to the $H_1$-$O_{17}$ hydrogen bond, $O_{17}$ of U forms an additional hydrogen bond with $H_{16}$ of the imine group of G after electron attachment. However, this is not observed for I:U base pair, which could be due to the absence of the amine group in hypoxanthine (I). For the I:U base pair anion, which has the second-highest RMSD value among wobble base pairs, the $O_6$–$H_{23}$ hydrogen bond length considerably increases by 0.603 Å. At the same time, the $H_{12}$–$O_{17}$ hydrogen bond in I:U strengthens as the bond compresses by 0.229 Å. This observation can also be attributed to the localization of the additional electron density on uracil, which enhances the hydrogen bond acceptor character of $O_{17}$ and makes the $N_{15}H_{23}$ group a less efficient hydrogen bond donor. The difference in the hydrogen bond lengths between the two anionic states of the base pairs does not follow any specific trend. However, it is strongly influenced by the nucleobase moiety on which the extra electron is localized. As evident from figure 2 and the RMSD values, this behavior could also be attributed to the varying extent to which the base pairs alter their geometry to accommodate the incoming electron. Similar to the differences in the hydrogen bond lengths between the two anionic structures, the geometric parameters of the individual nucleobases, especially the one on which the extra electron density is localized, are also affected by the formation of the valence-bound anion. The extra electron density is located on uracil (U) on G:U, and this is reflected in the structure by the elongation of C–N and C–C bond lengths in the heterocyclic ring.

Electron attachment to DNA can also result in proton transfer between the nucleobases. This process is of great interest because of the formation of nucleobase tautomers which can cause spontaneous mutations in DNA.[55–59] Sevilla and co-workers, based on their DFT studies, proposed that proton transfer is favored in I:C base pair anion.[60,61] They found that the proton transfer was driven by stronger inter-base interactions in the proton-transferred I:C anion.[60]

Several theoretical studies have been conducted on the electron attachment-induced proton transfer in WC G:C base pair in the gas phase.[60,62–65] Most of them, except the one by Honda et al.,[65] have not considered the role of dipole-bound anionic states in the process. Bowen and co-workers reported that the proton transfer in G:C anion may have a low-energy barrier.[64] They further proposed that proton transfer in G:C anion incorporated in DNA would have a small barrier of 2.39 kcal/mol based on the analysis of 9-methylguanine-1-methylcytosine dimer anion at B3LYP/6-31++G** level of theory. Schaefer and co-workers probed the proton transfer in 2′-deoxyribonucleoside dimer dG:dC upon electron attachment using B3LYP functional and DZP++ basis set and obtained a similar result.[62] Honda et al. considered the inter-base proton transfer in gas phase G:C as a two-step process, where the dipole-bound to valence-bound transition takes place initially and is followed by the proton transfer.[65] The barrier height (~2.7 kcal/mol) for proton transfer was determined with long-range corrected CAM-B3LYP functional and 6-311++G(2d,p) basis set with extra diffuse functions on hydrogen atoms. They have also performed ring polymer molecular dynamics simulations and found that the proton transfer occurs within ~10 ps.

Previous studies on the WC G:C base pair showed that the electron-accepting group in proton transfer is the nucleobase which houses the extra electron density.[60,64] I:C, I:A among the wobble base pairs exhibited proton transfer as the incoming electron got transferred from the dipole-bound to the valence-bound state. Cytosine in I:C, and Adenine in I:A act as the proton acceptors during the proton transfer. As one should expect, the extra electron density is localized on the proton acceptors in the valence-bound state. As shown in figure 1(b), electron and proton transfer both occur to cytosine. The electron is transferred to the $\pi^*$-type molecular orbital (MO) centered on cytosine from the dipole-bound state, and the proton transfer results in an N-H σ-bond formation on cytosine. Similarly, the electron and proton transfer take place to the adenine base in the I:A base pair.

The LEE attachment-induced proton transfer in I:A and I:C wobble base pairs could be a proton-coupled electron transfer (PCET) reaction. In the case of PCET reactions generally reported in the literature, the electron transfer occurs from a donor site to an acceptor site in the molecule or complex.[66] In those reactions, the electron density in the donor and acceptor sites before and after electron transfer, respectively, are localized on the nuclear framework of the system. However, the LEE attachment-induced proton transfer in I:C and I:A nucleobase pairs are different from such PCET reactions. This is because in the dipole-bound anion, from where the electron transfer occurs, is a diffuse state where the electron density is located away from the nuclear framework. The PCET in I:A and I:C wobble base pairs could proceed via two pathways. The first one is a two-step process where the formation of valence-bound anion from the dipole-bound state occurs initially, which is followed by the proton transfer. The second pathway involves a concerted electron-proton transfer (CEPT), where both electron and proton transfer occur simultaneously without the formation of an intermediate.[66] If the proton transfer reaction follows the first mechanism, there should be two barriers in the potential energy curve depicting the reaction – one between the dipole-bound state and the valence-bound states and the other between the valence-bound anion and the final product. We have performed a nudged elastic band (NEB) calculation to determine the minimum energy path (MEP) between the base pair anions in the valence-bound state and after proton transfer. If the second step has a potential energy barrier, it would indicate that the proton transfer is a two-step process with the valence-bound base pair anion as the intermediate. Conversely, if it is barrierless, it would suggest that it is a CEPT reaction. The MEP was recalculated at EA-EOM-DLPNO-CCSD/aug-cc-pVTZ(+5s5p4d) level of theory, using the intermediate geometries along the MEP obtained from NEB calculation and is given in figure 3. Clearly, the proton transfer from the I:C and I:A valence-bound anion is a barrier-less process, and therefore, it is a CEPT reaction. In comparison, the proton-coupled electron transfer in WC G:C anion is a two-step process as the MEP for the proton transfer (Figure S1) has a barrier, which is also consistent with existing theoretical studies.[65]

Although we did not observe inter-base proton transfer in I:U and G:U base pairs, we constructed proton transferred geometries for the two and determined the MEP for the reaction. The product geometries for the proton transfer reaction were obtained through constrained optimizations and subsequent frequency calculations were also carried out to ensure that they were not transition states. Although both I:U and G:U have two sites where proton transfer is possible, we have only considered a single proton transfer process. In case of I:U, proton transfer is possible between hypoxanthine $N_7H_{12}$ group and $O_{17}$ of uracil, and between uracil $N_{15}H_{23}$ and hypoxanthine $O_6$. We have considered the former and generated the MEP, as shown in figure 3(C). For G:U, we considered the proton transfer from $N_{27}H_{28}$ of uracil to $O_4$ of guanine. The MEP calculated for both I:U and G:U shows that the geometry formed after proton transfer is energetically unstable with respect to the reactant. Therefore, the proton transfer reactions we have considered are not likely to occur in I:U and G:U upon electron attachment.

Since the dipole-bound anionic state exists due to the charge-dipole interaction between the electron and the dipole moment vector, the extent to which they are vertically bound should depend on the magnitude of the dipole moment. Although the relationship between the dipole moment of the species and the corresponding VEA of the DNA subunit is often not quantitative.[54] The G:U (7.9 Debye) have the highest dipole moment among wobble base pairs and also have the highest vertical electron affinity (VEA) among them. VEA and adiabatic electron affinity (AEA) is calculated as follows:

$$VEA = E_{neutral}(\text{at optimized neutral geometry}) - E_{anion}(\text{at optimized neutral geometry}) \quad (1)$$

$$AEA = E_{neutral}(\text{at optimized neutral geometry}) - E_{anion}(\text{at optimized anion geometry}) \quad (2)$$

In comparison to the dipole moment and VEA reported by Tripathi et al. for WC base pairs,[39] G:U wobble base pair exhibited a larger dipole moment (7.9 Debye), and its dipole-bound anion is more strongly bound vertically.

The WC G:C and wobble I:C base pairs have almost the same structure except for the amino group of guanine, which is replaced by a hydrogen atom in hypoxanthine. Again, there are only two hydrogen bonds between hypoxanthine and cytosine in I:C compared to the three in WC G:C. The absence of the amino group might have resulted in the latter because of having a lower dipole moment (4.1 Debye) and VEA (68 meV) than that of the former, which has a dipole moment of 6.9 Debye and VEA of 99 meV at the same level of theory.[39] All the wobble base pairs have a higher dipole moment than WC A:T, and their dipole bound anions are more stable than A:T, except for I:A base pair which shows almost identical VEA as that of the A:T. Although a large dipole moment can cause stronger charge-dipole interaction in the dipole-bound anion, this trend is not always quantitative, as demonstrated by the VEA values obtained for I:C and I:U. The former has a lower dipole moment but a larger VEA than the latter (Table 1). Only G:U and I:C base pairs form adiabatically bound valence-bound anions among the four base pairs. The valence-bound anion of I:A is the most unstable in terms of AEA value, followed by I:U. Table 1 presents the VEA and AEA values determined at EA-EOM-DLPNO-CCSD/aug-cc-pVTZ(+5s5p4d) level. The AEA value for a particular base pair considerably rises when the size of the basis set is increased from aug-cc-pVDZ(+5s5p4d) to aug-cc-pVTZ(+5s5p4d). (See Table S2). We have extrapolated the values to complete basis set limit.

The difference in CBS level AEA values relative to those calculated using aug-cc-pVTZ(+5s5p4d) basis set for wobble base pairs are in the range of 56 meV (I:C) to 43 meV (I:U). This is presumably to the preferential stabilization of the valence-bound anion over the neutral caused by a larger basis set.

It has been previously shown that the dipole-bound state can act as a doorway to the formation of the valence-bound anion.[39,67–69] Our studies on bulk solvated nucleobases and WC base pairs showed that the doorway mechanism exists even in the condensed phase, where the water-bound state acts as the doorway.[40] To probe the mechanism of the dipole-bound to valence-bound transition, we have constructed a potential energy curve along a linear transit between the two states at EA-EOM-DLNPO-CCSD/aug-cc-pVTZ(+5s5p4d) level of theory. The intermediate geometries are given by the equation below:

$$R = (1-\lambda)R_{DB} + \lambda R_{VB} \qquad (3)$$

Where $R$ represents the geometrical parameter (bond length, bond angle, etc.) of the intermediate geometry which exists along the linear transit, $R_{DB}$ and $R_{VB}$ correspond to the value of the geometric parameter for dipole-bound and valence-bound geometries, respectively. $\lambda$ is a dimensionless parameter that is varied from zero to one to obtain the intermediate geometries. The value of $\lambda$ to be zero results in dipole-bound geometry, and if $\lambda = 1$ one gets the valence-bound anion geometry.

From figure 4a, it can be seen that the avoided crossing is present between the ground and the first excited state of the G:U anion. For $\lambda$ values closer to zero, only one bound state is observed, which is dipole-bound in nature and corresponds to the ground state of the anion. As we move along the adiabatic PEC, the first excited state of the anion (valence-bound in nature) also becomes bound just before the avoided crossing region. After the avoided crossing region, the nature of the anion's ground state becomes valence-bound, and that of the first excited state becomes dipole-bound. In the avoided crossing region, the nature of the wavefunction drastically changes as the two electronic states of the anion are energetically close. Here, the adiabatic picture is inadequate as the Born-Oppenheimer approximation breaks down since the two states become vibronically coupled.[70] One can treat such a situation in terms of diabatic states. In the diabatic picture, however, crossings between the PECs corresponding to two different electronic states are possible. Therefore, we have generated diabatic PECs from the adiabatic PEC by considering the valence-bound and dipole-bound anionic states as the basis.[71] We have used a simple avoided crossing model potential to calculate the coupling element between the two diabatic states given by the following expression:

$$V = \begin{pmatrix} V_1 & W \\ W & V_2 \end{pmatrix} \quad (4)$$

Here, $W$ is the coupling element. The diagonal elements $V_1$ and $V_2$ are chosen to be in the form of a fourth-degree polynomial (equation 2) as a function of the parameter $\lambda$ as follows:

$$V_i = a\lambda^4 + b\lambda^3 + c\lambda^2 + d\lambda + v_i^0 \quad (5)$$

The coupling constant calculated for G:U (18 meV) base pair is very small, which means the system is in a weak coupling limit. The coupling elements for the I:U and I:A base pairs are relatively higher but of the same order as reported for WC base pairs. I:C base pair showed the lowest coupling element (8 meV) among the four base pairs. The approximate rate for the electron transition from the dipole-bound state to the valence-bound state of the base pair can be calculated using Marcus theory. The expression for the rate constant based on Marcus theory of electron transfer is as follows:

$$k = \frac{2\pi}{\hbar}|W|^2 \sqrt{\frac{1}{4\pi k_B T \lambda_R}} e^{-(\lambda_R + \Delta G^0)^2 / 4\lambda_R k_B T} \quad (6)$$

Here, $\Delta G^0$ is the difference in free energy between the two states without considering the entropy contribution $(E_{VB} - E_{DB})$. $\lambda_R$ is the reorganization energy. The rate of dipole-bound state to valence-bound state transition for G:U has been found to be $3.7 \times 10^9$ s$^{-1}$. This shows that the dipole-bound state can act as a doorway to the formation of the valence-bound state since the latter will be populated quickly. However, this rate is around two orders less than the rate Tripathi et al. had reported for the WC G:C ($4.1 \times 10^{11}$ s$^{-1}$) base pair using the same level of theory.[39] It should be mentioned that they have taken the diagonal terms in the avoided crossing model potential as harmonic potential. We were able to obtain identical results for the

WC G:C base pair by using the quartic function to describe the $V_1$ and $V_2$ terms. We have also followed the same procedure for determining the rates for the remaining wobble base pairs. The I:C has the highest rate constant among the four wobble base pairs studied here, whereas I:A has the slowest rate with a rate constant of $5.6 \times 10^4$ s$^{-1}$. This slow rate can be attributed to the lack of stability of the valence-bound state of I:A. The rate constants depicted in Table 2 show that for all the wobble base pairs, the dipole-bound to valence-bound transition is slower than that of the WC G:C base pair. This indicates that the formation of a stable valence-bound state in non-canonical base pairs is kinetically less favorable relative to that in canonical WC G:C base pair.

Several studies have shown that the LEE attachment to DNA proceeds through the formation of a shape resonance state.[15,16,19,72] The location of this shape resonant state in DNA is a heavily debated matter, with the nucleobase[19] and the P=O of the phosphate group[24,72] arising as the two prominent candidates. Nevertheless, regardless of the location of the shape resonance in DNA, once it is formed, the electron gets transferred to the sugar-phosphate C–O bond resulting in its cleavage. Whereas the electron from the dipole-bound state can either get transferred to a stable valence-bound anionic state or lead to the bond break.[69,73] We have recently shown that the formation of a stable valence-bound state can be competitive to the process leading to the C–O bond cleavage.[69] Hence, the formation of a stable anion may cause suppression of sugar-phosphate bond cleavage and associated strand breaking. Since the rate of formation of the valence-bound state from the doorway dipole-bound state in wobble base pairs is slower than that in WC base pairs, the genetic materials containing wobble base pairs will be more prone to damage caused by LEE-induced attachment in gas phase.

**Conclusion**

In this study, we have shown that the electron attachment process in gas phase wobble base pairs happens through a doorway mechanism. The initial electron attachment leads to the generation of a dipole-bound anionic state, which acts as the doorway for the formation of a valence-bound state. The process of formation of the stable anion in non-WC base pairs via the doorway mechanism is kinetically less favored than that in WC G:C base pair. Therefore, biomolecules rich in non-WC base pairing may be more susceptible to damage due to LEE attachment. The addition of an extra electron leads to significant structural changes in the base pairs relative to the neutral geometry. Moreover, wobble I:C and I:A base pairs exhibited proton transfer during dipole-bound to valence-bound transition. The MEP for proton transfer obtained from NEB calculations shows that it is a concerted electron-proton transfer. On the contrary, the LEE attachment-induced proton transfer occurring in the WC G:C pair is a two-step process, where the first step involves the formation of the valence-bound anion from the dipole-bound one and the proton transfer takes place in the second step. The electron attachment to G:U and I:C leads to an adiabatically bound valence-bound state, whereas the corresponding state in the I:U and I:A are not adiabatically bound. However, in the condensed phase, due to favorable solvent-solute interactions, even I:U and I:A may form adiabatically bound anions, and the stability of the other base pair anions should also increase. Therefore, one may expect the rates of electron attachment to wobble base pairs to be higher in solutions. Work is in progress toward investigating the effect of explicit solvation on the LEE attachment process to the non-WC base pairs.

**Supporting Information**

Cartesian coordinates of the optimized neutral and anionic geometry of G:U, I:C, I:U and I:A Wobble base pairs, MEP for proton transfer in the VB anion of WC G:C base pair, basis set dependence of VEA and AEA values at EA-EOM-DLNPO-CCSD level is provided in the supporting information.

**Acknowledgments**

The authors acknowledge the support from the IIT Bombay, IIT Bombay Seed Grant project (Project no. RD/0517-IRCCSH0-040), DST-SERB (Project no. CRG/2018/001549), DST-Inspire Faculty Fellowship (Project no. DST/INSPIRE/04/2017/001730), Prime Minister's Research Fellowship for financial support, IIT Bombay super computational facility, and C-DAC supercomputing resources (PARAM Smriti, PARAM Brahma) for computational time.

**References**

(1) Pimblott, S. M.; LaVerne, J. A. On the Radiation Chemical Kinetics of the Precursor to the Hydrated Electron. *J. Phys. Chem. A* **1998**, *102* (17), 2967–2975.

(2) Kumar, A.; Becker, D.; Adhikary, A.; Sevilla, M. D. Reaction of Electrons with DNA: Radiation Damage to Radiosensitization. *Int. J. Mol. Sci.* **2019**, *20* (16), 3998.

(3) Alizadeh, E.; Sanche, L. Precursors of Solvated Electrons in Radiobiological Physics and Chemistry. *Chem. Rev.* **2012**, *112* (11), 5578–5602.

(4) Dong, Y.; Liao, H.; Gao, Y.; Cloutier, P.; Zheng, Y.; Sanche, L. Early Events in Radiobiology: Isolated and Cluster DNA Damage Induced by Initial Cations and Nonionizing Secondary Electrons. *J. Phys. Chem. Lett.* **2021**, *12* (1), 717–723.

(5) Gao, Y.; Zheng, Y.; Sanche, L. Low-Energy Electron Damage to Condensed-Phase DNA and Its Constituents. *Int. J. Mol. Sci.* **2021**, *22* (15) 7879.

(6) Woldhuis, J.; Verberne, J. B.; Lafleur, M. V. M.; Retèl, J.; Blok, J.; Loman, H. γ-Rays Inactivate ΦX174 DNA in Frozen Anoxic Solutions at −20°C Mainly by Reactions of Dry Electrons. *Int. J. Rad. Biol.* **1984**, *46* (4), 329–330.

(7) Boudaïffa, B.; Cloutier, P.; Hunting, D.; Huels, M. A.; Sanche, L. Resonant Formation of DNA Strand Breaks by Low-Energy (3 to 20 EV) Electrons. *Science* **2000**, *287* (5458), 1658–1660.

(8) Abdoul-Carime, H.; Gohlke, S.; Fischbach, E.; Scheike, J.; Illenberger, E. Thymine Excision from DNA by Subexcitation Electrons. *Chem. Phys. Lett.* **2004**, *387* (4–6), 267–270.

(9) Zheng, Y.; Cloutier, P.; Hunting, D. J.; Sanche, L.; Wagner, J. R. Chemical Basis of DNA Sugar-Phosphate Cleavage by Low-Energy Electrons. *J. Am. Chem. Soc.* **2005**, *127* (47), 16592–16598.

(10) Sanche, L. Low Energy Electron-Driven Damage in Biomolecules. *Eur. Phys. J. D* **2005**, *35* (2), 367–390.

(11) Zheng, Y.; Cloutier, P.; Hunting, D. J.; Wagner, J. R.; Sanche, L. Phosphodiester and N-Glycosidic Bond Cleavage in DNA Induced by 4-15 EV Electrons. *J. Chem. Phys.* **2006**, *124* (6) 064710.

(12) Pan, X.; Cloutier, P.; Hunting, D.; Sanche, L. Dissociative Electron Attachment to DNA. *Phys. Rev. Lett.* **2003**, *90* (20), 208102.

(13) Khorsandgolchin, G.; Sanche, L.; Cloutier, P.; Wagner, J. R. Strand Breaks Induced by Very Low Energy Electrons: Product Analysis and Mechanistic Insight into the Reaction with TpT. *J. Am. Chem. Soc.* **2019**, *141* (26), 10315–10323.

(14) Kumari, B.; Huwaidi, A.; Robert, G.; Cloutier, P.; Bass, A. D.; Sanche, L.; Wagner, J. R. Shape Resonances in DNA: Nucleobase Release, Reduction, and Dideoxynucleoside Products Induced by 1.3 to 2.3 EV Electrons. *J. Phys. Chem. B* **2022**, *126* (28), 5175–5184.

(15) Barrios, R.; Skurski, P.; Simons, J. Mechanism for Damage to DNA by Low-Energy Electrons. *J. Phys. Chem. B* **2002**, *106* (33), 7991–7994.

(16) Berdys, J.; Anusiewicz, I.; Skurski, P.; Simons, J. Theoretical Study of Damage to DNA by 0.2-1.5 EV Electrons Attached to Cytosine. *J. Phys. Chem. A* **2004**, *108* (15), 2999–3005.

(17) Berdys, J.; Anusiewicz, I.; Skurski, P.; Simons, J. Damage to Model DNA Fragments from Very Low-Energy (<1 EV) Electrons. *J. Am. Chem. Soc.* **2004**, *126* (20), 6441–6447.

(18) Anusiewicz, I.; Berdys, J.; Sobczyk, M.; Skurski, P.; Simons, J. Effects of Base π-Stacking on Damage to DNA by Low-Energy Electrons. *J. Phys. Chem. A* **2004**, *108* (51), 11381–11387.

(19) Simons, J. How Do Low-Energy (0.1-2 EV) Electrons Cause DNA-Strand Breaks? *Accounts of Chemical Research* **2006**, *39* (10), 772–779.
(20) Gu, J.; Wang, J.; Leszczynski, J. Electron Attachment-Induced DNA Single Strand Breaks: C3′- O3′ σ-Bond Breaking of Pyrimidine Nucleotides Predominates. *J. Am. Chem. Soc.* **2006**, *128* (29), 9322–9323.
(21) Bao, X.; Wang, J.; Gu, J.; Leszczynski, J. DNA Strand Breaks Induced by Near-Zero-Electronvolt Electron Attachment to Pyrimidine Nucleotides. *Proceedings of the National Academy of Sciences of the United States of America* **2006**, *103* (15), 5658–5663.
(22) Gu, J.; Xie, Y.; Schaefer, H. F. Near 0 EV Electrons Attach to Nucleotides. *J. Am. Chem. Soc.* **2006**, *128* (4), 1250–1252.
(23) Bhaskaran, R.; Sarma, M. The Role of the Shape Resonance State in Low Energy Electron Induced Single Strand Break in 2′-Deoxycytidine-5′-Monophosphate. *Phys. Chem. Chem. Phys.* **2015**, *17* (23), 15250–15257.
(24) Bhaskaran, R.; Sarma, M. Low-Energy Electron Interaction with the Phosphate Group in DNA Molecule and the Characteristics of Single-Strand Break Pathways. *J. Phys. Chem. A* **2015**, *119* (40), 10130–10136.
(25) Colson, A. O.; Besler, B.; Sevilla, M. D. Ab Initio Molecular Orbital Calculations on DNA Base Pair Radical Ions: Effect of Base Pairing on Proton-Transfer Energies, Electron Affinities, and Ionization Potentials. *J. Phys. Chem.* **1992**, *96* (24), 9787–9794.
(26) Al-Jihad, I.; Smets, J.; Adamowicz, L. Covalent Anion of the Canonical Adenine−Thymine Base Pair. Ab Initio Study. *J. Phys. Chem. A* **2000**, *104* (13), 2994–2998.
(27) Smets, J.; Jalbout, A. F.; Adamowicz, L. Anions of the Hydrogen-Bonded Guanine–Cytosine Dimer – Theoretical Study. *Chem. Phys. Lett.* **2001**, *342* (3), 342–346.
(28) Stanton, J. F.; Gauss, J. A Discussion of Some Problems Associated with the Quantum Mechanical Treatment of Open-Shell Molecules. In *Advances in Chemical Physics*; John Wiley & Sons, Ltd, 2003; pp 101–146.
(29) Richardson, N. A.; Wesolowski, S. S.; Schaefer, H. F. Electron Affinity of the Guanine−Cytosine Base Pair and Structural Perturbations upon Anion Formation. *J. Am. Chem. Soc.* **2002**, *124* (34), 10163–10170.
(30) Richardson, N. A.; Wesolowski, S. S.; Schaefer, H. F. The Adenine−Thymine Base Pair Radical Anion: Adding an Electron Results in a Major Structural Change. *J. Phys. Chem. B* **2003**, *107* (3), 848–853.
(31) Kim, M.-C.; Sim, E.; Burke, K. Communication: Avoiding Unbound Anions in Density Functional Calculations. *J. Chem. Phys.* **2011**, *134* (17), 171103.
(32) ROWE, D. J. Equations-of-Motion Method and the Extended Shell Model. *Rev. Mod. Phys.* **1968**, *40* (1), 153–166.
(33) Voora, V. K.; Kairalapova, A.; Sommerfeld, T.; Jordan, K. D. Theoretical Approaches for Treating Non-Valence Correlation-Bound Anions. *J. Chem. Phys.* **2017**, *147* (21), 214114.
(34) Nooijen, M.; Bartlett, R. J. Equation of Motion Coupled Cluster Method for Electron Attachment. *J. Chem. Phys.* **1995**, *102* (9), 3629–3647.
(35) Sekino, H.; Bartlett, R. J. A Linear Response, Coupled-Cluster Theory for Excitation Energy. *Int. J. Quantum Chem.* **1984**, *26* (S18), 255–265.
(36) Krylov, A. I. Equation-of-Motion Coupled-Cluster Methods for Open-Shell and Electronically Excited Species: The Hitchhiker's Guide to Fock Space. *Annu. Rev. Phys. Chem.* **2008**, *59* (1), 433–462.
(37) Riplinger, C.; Neese, F. An Efficient and near Linear Scaling Pair Natural Orbital Based Local Coupled Cluster Method. *J. Chem. Phys.* **2013**, *138* (3), 034106.

(38) Dutta, A. K.; Saitow, M.; Demoulin, B.; Neese, F.; Izsák, R. A Domain-Based Local Pair Natural Orbital Implementation of the Equation of Motion Coupled Cluster Method for Electron Attached States. *J. Chem. Phys.* **2019**, *150* (16), 164123.
(39) Tripathi, D.; Dutta, A. K. Electron Attachment to DNA Base Pairs: An Interplay of Dipole- And Valence-Bound States. *J. Phys. Chem. A* **2019**, *123* (46), 10131–10138.
(40) Ranga, S.; Mukherjee, M.; Dutta, A. K. Interactions of Solvated Electrons with Nucleobases: The Effect of Base Pairing. *ChemPhysChem* **2020**, *21* (10), 1019–1027.
(41) Gu, J.; Leszczynski, J.; Schaefer, H. F. Interactions of Electrons with Bare and Hydrated Biomolecules: From Nucleic Acid Bases to DNA Segments. *Chem. Rev.* **2012**, *112* (11), 5603–5640.
(42) Leontis, N. B.; Westhof, E. Geometric Nomenclature and Classification of RNA Base Pairs. *RNA* **2001**, *7* (4), 499–512.
(43) Nikolova, E. N.; Zhou, H.; Gottardo, F. L.; Alvey, H. S.; Kimsey, I. J.; Al-Hashimi, H. M. A Historical Account of Hoogsteen Base-Pairs in Duplex DNA. *Biopolymers* **2013**, *99* (12), 955–968.
(44) Hou, Y.-M.; Schimmel, P. A Simple Structural Feature Is a Major Determinant of the Identity of a Transfer RNA. *Nature* **1988**, *333* (6169), 140–145.
(45) Weigend, F.; Häser, M. RI-MP2: First Derivatives and Global Consistency. *Theor. Chem. Acc.* **1997**, *97* (1), 331–340.
(46) Weigend, F.; Häser, M.; Patzelt, H.; Ahlrichs, R. RI-MP2: Optimized Auxiliary Basis Sets and Demonstration of Efficiency. *Chem. Phys. Lett.* **1998**, *294* (1), 143–152.
(47) Weigend, F.; Ahlrichs, R. Balanced Basis Sets of Split Valence, Triple Zeta Valence and Quadruple Zeta Valence Quality for H to Rn: Design and Assessment of Accuracy. *Phys. Chem. Chem. Phys.* **2005**, *7* (18), 3297–3305.
(48) Neese, F. The ORCA Program System. *WIREs Comput. Mol. Sci.* **2012**, *2* (1), 73–78.
(49) Neese, F. Software Update: The ORCA Program System, Version 4.0. *WIREs Comput. Mol. Sci.* **2018**, *8* (1), e1327.
(50) Neese, F.; Wennmohs, F.; Becker, U.; Riplinger, C. The ORCA Quantum Chemistry Program Package. *J. Chem. Phys.* **2020**, *152* (22), 224108.
(51) Dunning, T. H. Gaussian Basis Sets for Use in Correlated Molecular Calculations. I. The Atoms Boron through Neon and Hydrogen. *J. Chem. Phys.* **1989**, *90* (2), 1007–1023.
(52) Ásgeirsson, V.; Birgisson, B. O.; Bjornsson, R.; Becker, U.; Neese, F.; Riplinger, C.; Jónsson, H. Nudged Elastic Band Method for Molecular Reactions Using Energy-Weighted Springs Combined with Eigenvector Following. *J. Chem. Theory Comput.* **2021**, *17* (8), 4929–4945.
(53) Gutsev, G. L.; Bartlett, R. J. A Theoretical Study of the Valence- and Dipole-bound States of the Nitromethane Anion. *J. Chem. Phys.* **1996**, *105* (19), 8785–8792.
(54) Tripathi, D.; Dutta, A. K. Bound Anionic States of DNA and RNA Nucleobases: An EOM-CCSD Investigation. *Int. J. Quantum Chem.* **2019**, *119* (9), e25875.
(55) WATSON, J. D.; CRICK, F. H. C. Genetical Implications of the Structure of Deoxyribonucleic Acid. *Nature* **1953**, *171* (4361), 964–967.
(56) Löwdin, P.-O. Proton Tunneling in DNA and Its Biological Implications. *Rev. Mod. Phys.* **1963**, *35* (3), 724–732.
(57) Topal, M. D.; Fresco, J. R. Complementary Base Pairing and the Origin of Substitution Mutations. *Nature* **1976**, *263* (5575), 285–289.
(58) Rüterjans, H.; Kaun, E.; Hull, W. E.; Limbach, H. H. Evidence for Tautornerisrn in Nucleic Acid Base Pairs. 1 H NMR Study of 15 N Labeled TRNA. *Nucleic Acids Res.* **1982**, *10* (21), 7027–7039.
(59) Purrello, R.; Molina, M.; Wang, Y.; Smulevich, G.; Fresco, J. R.; Spiro, T. G.; Fossella, J. Keto-Iminol Tautomerism of Protonated Cytidine Monophosphate Characterized by

Ultraviolet Resonance Raman Spectroscopy: Implications of C+ Iminol Tautomer for Base Mispairing. *J. Am. Chem. Soc.* **1993**, *115* (2), 760–767.
(60) Li, X.; Cai, Z.; Sevilla, M. D. Investigation of Proton Transfer within DNA Base Pair Anion and Cation Radicals by Density Functional Theory (DFT). *J. Phys. Chem. B* **2001**, *105* (41), 10115–10123.
(61) Li, X.; Cai, Z.; Sevilla, M. D. Energetics of the Radical Ions of the AT and AU Base Pairs: A Density Functional Theory (DFT) Study. *J. Phys. Chem. A* **2002**, *106* (40), 9345–9351.
(62) Gu, J.; Xie, Y.; Schaefer, H. F. Electron Attachment Induced Proton Transfer in a DNA Nucleoside Pair: 2′-Deoxyguanosine-2′-Deoxycytidine. *J. Chem. Phys.* **2007**, *127* (15), 155107.
(63) Szyperska, A.; Rak, J.; Leszczynski, J.; Li, X.; Ko, Y. J.; Wang, H.; Bowen, K. H. Valence Anions of 9-Methylguanine−1-Methylcytosine Complexes. Computational and Photoelectron Spectroscopy Studies. *J. Am. Chem. Soc.* **2009**, *131* (7), 2663–2669.
(64) Szyperska, A.; Rak, J.; Leszczynski, J.; Li, X.; Ko, Y. J.; Wang, H.; Bowen, K. H. Low-Energy-Barrier Proton Transfer Induced by Electron Attachment to the Guanine⋯Cytosine Base Pair. *ChemPhysChem* **2010**, *11* (4), 880–888.
(65) Honda, T.; Minoshima, Y.; Yokoi, Y.; Takayanagi, T.; Shiga, M. Semiclassical Dynamics of Electron Attachment to Guanine–Cytosine Base Pair. *Chem. Phys. Lett.* **2015**, *625*, 174–178.
(66) Tyburski, R.; Liu, T.; Glover, S. D.; Hammarström, L. Proton-Coupled Electron Transfer Guidelines, Fair and Square. *J. Am. Chem. Soc.* **2021**, *143* (2), 560–576.
(67) Mukherjee, M.; Tripathi, D.; Dutta, A. K. Water Mediated Electron Attachment to Nucleobases: Surface-Bound vs Bulk Solvated Electrons. *J. Chem. Phys.* **2020**, *153* (4), 044305.
(68) Verma, P.; Ghosh, D.; Dutta, A. K. Electron Attachment to Cytosine: The Role of Water. *J. Phys. Chem. A* **2021**, *125* (22), 4683–4694.
(69) Narayanan S J, J.; Tripathi, D.; Dutta, A. K. Doorway Mechanism for Electron Attachment Induced DNA Strand Breaks. *J. Phys. Chem. Lett.* **2021**, *12* (42), 10380–10387.
(70) Köuppel, H.; Domcke, W.; Cederbaum, L. S. Multimode Molecular Dynamics Beyond the Born-Oppenheimer Approximation. In *Adv. Chem. Phys.*; John Wiley & Sons, Ltd, 1984; pp 59–246.
(71) Sommerfeld, T. Intramolecular Electron Transfer from Dipole-Bound to Valence Orbitals: Uracil and 5-Chlorouracil. *J. Phys. Chem. A* **2004**, *108* (42), 9150–9154.
(72) Kumar, A.; Sevilla, M. D. The Role of Πσ* Excited States in Electron-Induced DNA Strand Break Formation: A Time-Dependent Density Functional Theory Study. *J. Am. Chem. Soc.* **2008**, *130* (7), 2130–2131.
(73) Kang, D. H.; Kim, J.; Eun, H. J.; Kim, S. K. Experimental Observation of the Resonant Doorways to Anion Chemistry: Dynamic Role of Dipole-Bound Feshbach Resonances in Dissociative Electron Attachment. *J. Am. Chem. Soc.* **2022**, *144* (35), 16077–16085.

**Table 1.** RMSD, dipole moment, vertical electron affinity (VEA), and adiabatic electron affinity (AEA) values of the wobble base pairs calculated at EA-EOM-DLNPO-CCSD/aug-cc-pVTZ(+5s5p4d) level of theory.

| base pair | RMSD (Å) | Dipole moment (Debye) | VEA (meV) | AEA (meV) |
|---|---|---|---|---|
| G:U | 1.333 | 7.9 | 107 | 53 |
| I:C | 0.237 | 4.1 | 68 | 334 |
| I:U | 0.647 | 6.1 | 62 | -176 |
| I:A | 0.305 | 3.5 | 3 | -355 |

**Table 2.** The coupling element and the rate constant values of the wobble and Hoogsteen base pairs determined from PECs calculated at EOM-DLPNO-CCSD/aug-cc-pVTZ(+5s5p4d) level of theory.

| Base pair | $W$ (meV) | rate constant |
|---|---|---|
| G:U | 18 | $3.7 \times 10^9$ |
| I:C | 8 | $4.5 \times 10^{10}$ |
| I:U | 26 | $1.3 \times 10^6$ |
| I:A | 26 | $5.6 \times 10^4$ |

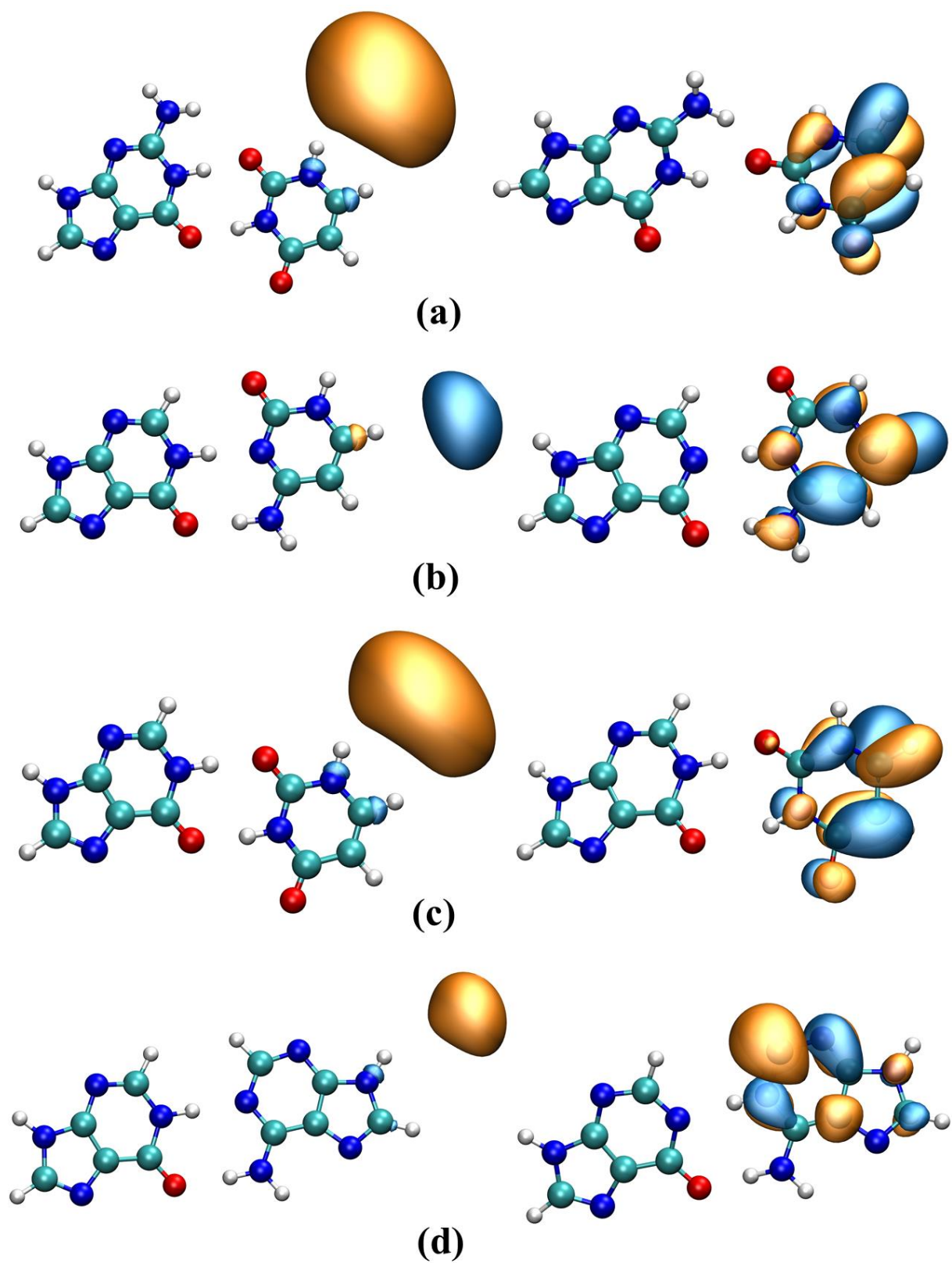

**Figure 1.** The molecular orbitals corresponding to the dominant transition in the EA-EOM-DLPNO-CCSD method for dipole-bound state (left) and valence-bound state (right) of wobble (a) G:U, (b) I:C, (c) I:U, (d) I:A base pairs.

**Figure 2.** Geometries of neutral (right) and valence-bound anion of (left) wobble (a) G:U, (b) I:C, (c) I:U, and (d) I:A wobble base pairs.

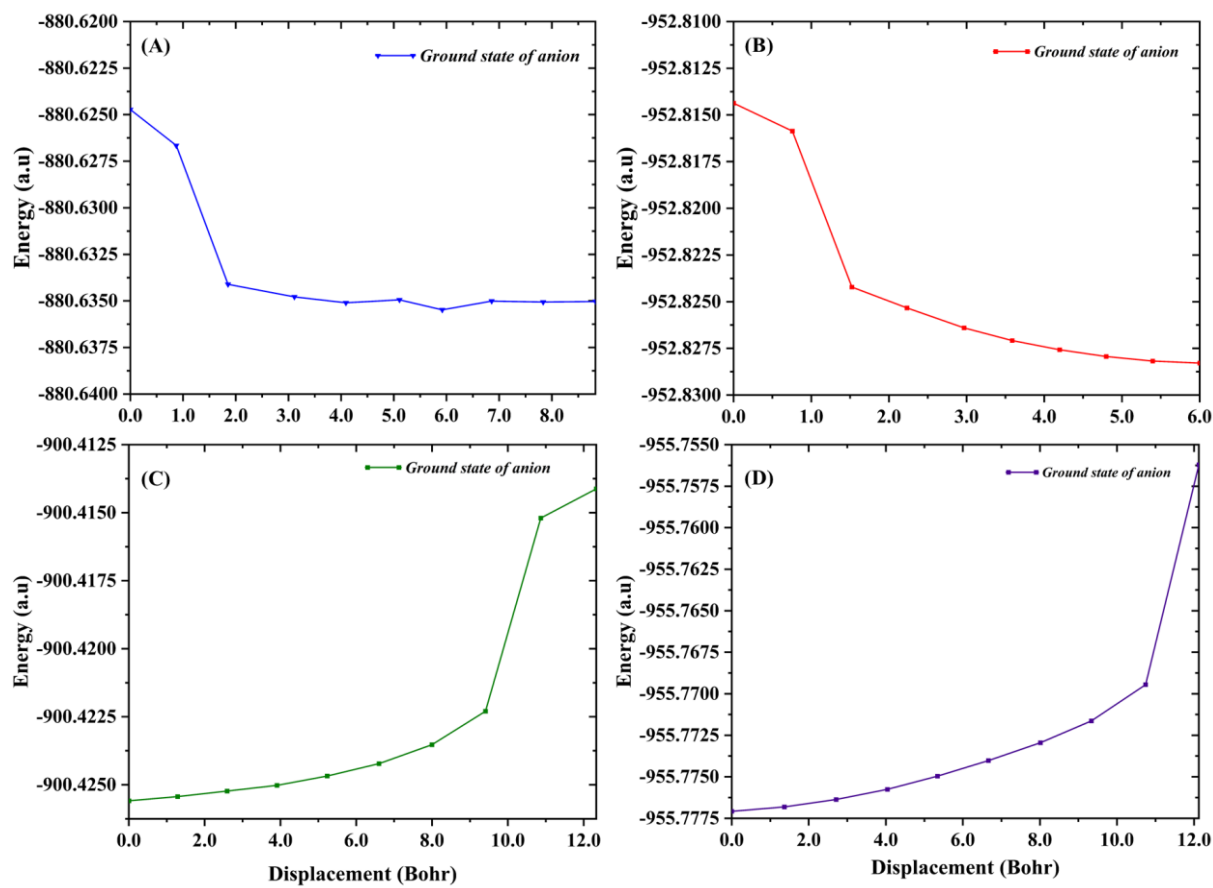

**Figure 3.** MEP of proton transfer reaction from the valence-bound anion of (A) I:C, (B) I:A, (C) I:U, and (D) G:U base pairs calculated at EA-EOM-DLPNO-CCSD/aug-cc-pVTZ(+5s5p4d) level of theory. The intermediate geometries were obtained from an NEB calculation conducted at the level of theory used for optimization.

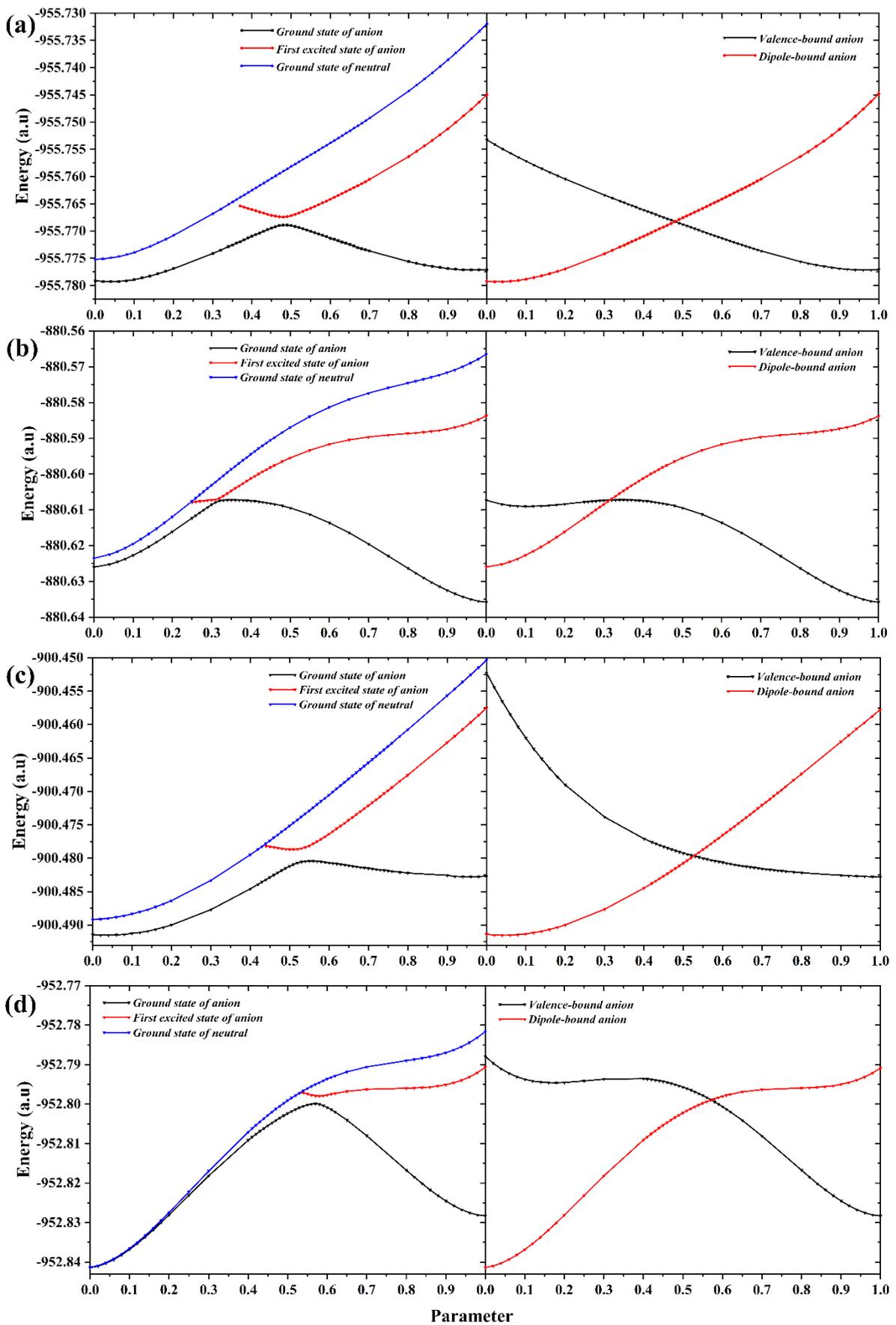

**Figure 4.** Adiabatic (left) and diabatic (diabatic) PEC corresponding to the dipole-bound to valence-bound transition for wobble (a) G:U, (b) I:C, (a) I:U and (b) I:A base pairs.